# A New High Voltage 4H-SiC Lateral Dual Sidewall Schottky (LDSS) Rectifier: Theoretical Investigation and Analysis

M. Jagadesh Kumar[1], *Senior Member, IEEE* and C. Linga Reddy[2], *Student Member, IEEE*

Department of Electrical Engineering
Indian Institute of Technology, Delhi
Hauz Khas, New Delhi – 110 016, INDIA

Email: mamidala@ieee.org    FAX: 91-11-658 1264

---

[1]Author for communication.

# Abstract

In this paper, we report a new 4H-SiC Lateral Dual Sidewall Schottky (LDSS) rectifier on a highly doped drift layer consisting of a high-barrier sidewall Schottky contact on top of the low-barrier Schottky contact. Using two-dimensional device simulation, the performance of the proposed device has been evaluated in detail by comparing its characteristics with those of the compatible Lateral Conventional Schottky (LCS) and Lateral Trench Sidewall Schottky (LTSS) rectifiers on 4H-SiC. From our simulation results, it is observed that the proposed LDSS rectifier acts as a low-barrier LTSS rectifier under forward bias conditions and as a high-barrier LTSS rectifier under reverse bias conditions making it an ideal rectifier. The LDSS rectifier exhibits an on/off current ratio (at 1 V / -500 V) of $5.5 \times 10^7$ for an epitaxial layer doping of $1 \times 10^{17}$ /cm$^3$. Further, the proposed LDSS structure exhibits a very sharp breakdown similar to that of a PiN diode in spite of using only Schottky junctions in the structure. We have analyzed the reasons for the improved performance of the LDSS.

# 1. INTRODUCTION

SiC Schottky rectifiers are now well known for their advantages in high temperature, high speed and high voltage applications [1]. Although several vertical SiC Schottky rectifiers have been reported in literature [2], lateral Schottky rectifiers are increasingly becoming important because of their utility in power ICs [3-5]. Recently, Kumar and Singh [5] have demonstrated that Schottky rectifiers with low reverse leakage current and PiN diode like sharp reverse breakdown can be achieved using the sidewall Schottky contact of a trench filled with a metal. Selecting a metal for this Lateral Trench Sidewall Schottky (LTSS) rectifier with a suitable work-function plays a vital role in deciding its forward and reverse characteristics. If a low-barrier metal is used for the sidewall Schottky contact in the LTSS rectifier, it will result in a low forward voltage drop but the reverse leakage current will be large and vice-versa for a high-barrier Schottky contact. However, it would be ideal if the Schottky rectifier has both low forward voltage drop as well as low reverse leakage current. This is impossible to obtain in a LTSS rectifier without making a trade-off between forward voltage drop and the reverse leakage current. In order to overcome this, for the first time, we propose a Lateral Dual Sidewall Schottky (LDSS) rectifier in which the low-barrier sidewall Schottky conducts during forward bias and the high-barrier sidewall Schottky conducts during the reverse bias. Based on simulation results, we demonstrate that the forward characteristics of the proposed LDSS structure are close to that of the low-barrier LTSS rectifier and the reverse characteristics are close to that of the high-barrier LTSS rectifier resulting in a near ideal Schottky rectifier. The reverse breakdown of the LDSS structure is also very sharp and is similar to that of a PiN diode. In the following sections, we report the results on this unique and highly desired I-V characteristics of the proposed 4H-SiC LDSS rectifier and analyze the reasons for the improved performance. It may be pointed out that although we have chosen SiC as the material for studying the LDSS structure, our basic idea should work equally well for other materials such as silicon whose technology is more matured than that of SiC.

## 2. LDSS STRUCTURE AND FABRICATION STEPS

Schematic cross-sectional view of the 4H-SiC LDSS rectifier implemented in the 2-dimensional device simulator MEDICI [6] is shown in Fig. 1. The anode of the LDSS rectifier consists of a high barrier Schottky (HBS) metal (Nickel with $\phi_{BH}$ = 1.50 eV) on top of the low barrier Schottky (LBS) metal (Titanium with $\phi_{BL}$ = 0.85 eV). These two metals are commonly used in SiC Schottky rectifiers [1]. Electric field crowding at the trench edges is reduced by using a simple field plate termination. The ohmic cathode contact is taken from the $N^+$ region. The dual sidewall Schottky contact can be created in the 4H-SiC LDSS structure as shown in Fig. 2. We start with a semi-insulating 4H-SiC substrate with a 2.0 μm n-type 4H-SiC epitaxial layer doped at $N_D$ = 1×10$^{17}$ cm$^{-3}$. The low-barrier Titanium metal of desired thickness can be first deposited in the trench as shown in Fig. 2(a). Using another mask and selective etch process, a second trench is made resulting in the low-barrier Schottky contact as shown in Fig. 2(b). Following this step, Nickel can be deposited to form the high-barrier Schottky contact as illustrated in Fig. 2(c). The final structure will be as shown in Fig. 1. The 4H-SiC LDSS rectifier is compared with the compatible 4H-SiC Lateral Conventional Schottky (LCS) and the Lateral Trench Sidewall Schottky (LTSS) rectifiers whose structures are similar to those reported in [5].

# 3. OPTIMIZATION OF LDSS DEVICE PARAMETERS

For optimizing the device parameters, we have first fixed the thickness ($t$), doping ($N_D$) and length (L) of the drift region by considering the series resistance and breakdown voltage. We have varied the low-barrier Schottky trench depth ($d_l$) and high-barrier Schottky trench depth ($d_h$) to get the optimum I-V behaviour. The performance of the proposed device (LDSS) in terms of forward voltage drop (@100 A/cm$^2$) and reverse current density (@500 V) are tabulated in Table-1 for various $d_l$ and $d_h$. From this data, we extracted the optimum value of the low-barrier Schottky trench depth ($d_l$) as 0.25 μm and the high-barrier Schottky trench depth ($d_h$) as 1.75 μm. To find the optimum field oxide thickness ($t_{ox}$), we have extracted the breakdown voltage and the reverse current density (@500 V) for various values of field oxide thickness as shown in Table-2. From this data, we extracted the optimum field oxide thickness ($t_{ox}$) to be 0.4 μm for the highest breakdown voltage. Field plate length ($L_{FP}$), is chosen to be 4.50 μm such that the breakdown voltage of the proposed device is maximum for the above optimum parameters. The final optimized device parameters used in the simulation for the 4H-SiC LDSS rectifier are given in Table-3.

# 4. SIMULATION RESULTS AND DISCUSSION

*A. Barrier height lowering model*

Simulation of SiC Schottky rectifiers is not easy because the measured reverse current density for 4H-SiC Schottky diodes has been reported to be much higher than predicted by thermionic emission theory and depends strongly on the applied voltage [7]. This discrepancy between measured and estimated reverse leakage current is due to the complex dependence of barrier height on image forces, surface inhomogeneities [8], depletion region generation [2], carrier tunneling [9] and also the barrier height fluctuations. Incorporating all these current flow mechanisms in the thermionic emission model is very difficult primarily because the fundamental physics taking place at the Schottky interface is not well understood. To overcome this problem, Singh and Kumar [1] have proposed a simple empirical model for the barrier height lowering ($\Delta\phi_B$) based on experimental results and have shown its application in two-dimensional simulation of 4H-SiC Schottky rectifiers to accurately predict both the forward and reverse characteristics. The Singh-Kumar barrier height lowering model [1] for 4H-SiC Schottky structures can be expressed as

$$\Delta\phi_B = a[E_{av}]^{1/2} + b \qquad (1)$$

where $E_{av}$ is the average electric field (V/cm) at the Schottky contact and a and b are constants. These constants have been reported to be a = $3.63 \times 10^{-4}$ $V^{1/2}cm^{1/2}$ and b = -0.034 V for Ti and a = $1.54 \times 10^{-4}$ $V^{1/2}cm^{1/2}$ and b = 0.638 V for Ni, the metals used in our study. We have implemented the above Singh-Kumar barrier lowering model for 4H-SiC in our 2-D numerical simulation and evaluated the performance of the proposed structure as discussed below.

*B. Forward and Reverse Characteristics*

Fig. 3 shows the simulated forward and reverse characteristics of the 4H-SiC low-barrier and high-barrier LCS, low-barrier and high-barrier LTSS and the proposed LDSS rectifiers. From Fig. 3(a), we observe that the forward characteristic of the LDSS rectifier is close to that of low-barrier LCS or LTSS rectifier as most of the current flows through low-barrier Schottky contact during forward bias as shown by the current vectors in Fig.4(a). The forward voltage drop of the proposed LDSS structure is approximately 0.55 V at a current density of 100

A/cm$^2$. From Fig. 3(b), we note that the reverse leakage current of the LDSS rectifier is very low and is close to that of the high-barrier LCS and LTSS rectifiers. This is because during the reverse bias the low-barrier Schottky is shielded by the depletion region and most of the reverse current flows through the high-barrier Schottky contact. This can be seen from the reverse current vectors of the LDSS rectifier shown in Fig. 4(b). We also note that the reverse current of both the low-barrier and high-barrier LCS rectifiers increases very rapidly with increasing reverse voltage resulting in an extremely soft breakdown at 230 V. However, the breakdown voltage of the LDSS rectifier is very large at ~ 1000 V (more than four times the breakdown voltage of LCS rectifiers.). This is because the breakdown of the LDSS structure takes place below the right edge of the field plate (away from the Schottky contact) [5]. To understand this behavior, the electric field is plotted in Fig. 5 along the horizontal line at the field-oxide/4H-SiC interface of low-barrier and high-barrier LCS, low-barrier and high-barrier LTSS and LDSS rectifiers near the breakdown voltage of the LCS rectifier (230 V). We note from Fig. 5 that in the case of low-barrier and high-barrier LCS rectifiers, the peak electric field is large and occurs at the Schottky junction resulting in a large barrier lowering as shown in Fig. 6. This leads to a soft and low breakdown in the case of low-barrier and high-barrier LCS rectifiers. But, the proposed LDSS structure exhibits a very sharp breakdown similar to that of a PiN diode in spite of using only Schottky junctions in the structure. This is due to the reduced electric field at the Schottky contact which in turn results in a significantly diminished barrier lowering ($\Delta\phi_B$) for the LDSS structure as shown in Fig. 6.

A useful figure of merit for rectifiers which combines the current carrying capability under forward bias and blocking capability under reverse bias, is the on/off current ratio [7] defined at fixed forward (1 V) and reverse biases (–500 V). Taking the *J-V* characteristics into consideration, the calculated on/off current ratio for the LDSS rectifier is of about $5.5 \times 10^7$ at 1 V/-500 V for an epitaxial layer doping of $1 \times 10^{17}/cm^3$, which is same as the on/off current ratio obtained by considering the forward characteristic of the low-barrier LTSS and reverse characteristic of the high-barrier LTSS. In other words, the LDSS structure combines the benefits of the low-barrier LTSS and high barrier LTSS rectifiers resulting in an ideal Schottky rectifier.

## 5. CONCLUSION

A novel high-voltage 4H-SiC Lateral Dual Sidewall Schottky (LDSS) rectifier has been presented. Using 2-dimensional simulation, we have demonstrated that the forward characteristic of the proposed LDSS rectifier is close to that of a low-barrier LTSS rectifier and its reverse characteristic is close to that of a high-barrier LTSS rectifier, resulting in an on/off current ratio of $5.5 \times 10^7$. An interesting feature of the proposed LDSS rectifier is that it exhibits a sharp breakdown similar to that of a PiN diode in spite of using only Schottky junctions in the LDSS structure. The combined low forward voltage drop, low reverse leakage current and excellent reverse blocking capability make the proposed LDSS rectifier attractive for use in low-loss, high-voltage and high-speed power IC applications.

## ACKNOWLEDGMENT

We would like to thank Council of Scientific and Industrial Research (CSIR), Government of India, for supporting Mr. Linga Reddy in his research.

# Figure captions

Fig. 1  Cross-sectional view of the 4H-SiC Lateral Dual Sidewall Schottky (LDSS).

Fig. 2  Steps for creating dual metal sidewall Schottky contacts in 4H-SiC LDSS rectifier.

Fig. 3 (a) Forward conduction and (b) Reverse blocking characteristics of the low-barrier and high-barrier LCS, low-barrier and high-barrier LTSS and LDSS rectifiers.

Fig. 4 (a) Current flow vectors in the LDSS rectifier at a forward current density of 100A/cm$^2$ (b) Current flow vectors in the LDSS rectifier at a reverse bias voltage of 500V.

Fig. 5 Electric field variation along the horizontal line at the field-oxide / 4H-SiC interface of low-barrier and high-barrier LCS, low-barrier and high-barrier LTSS and LDSS rectifiers near the breakdown voltage of low-barrier and high-barrier LCS rectifiers (230 V).

Fig. 6 Barrier height lowering as a function of reverse bias voltage for the low-barrier and high-barrier LCS, low-barrier and high-barrier LTSS and LDSS rectifiers.

Table 1: Forward voltage drop (@100 A/cm$^2$) and Reverse current density (@500 V) for various $d_l$ and $d_h$.

| $d_l$ (μm) | $d_h$ (μm) | Forward voltage drop @100 A/cm$^2$ | Reverse current density @500 V |
|---|---|---|---|
| 0.00 | 2.00 | 1.21 | 6.00E-05 |
| 0.25 | 1.75 | 0.55 | 7.15E-05 |
| 0.50 | 1.50 | 0.55 | 2.52E-04 |
| 0.75 | 1.25 | 0.55 | 5.91E-03 |
| 1.00 | 1.00 | 0.55 | 4.25E-02 |
| 1.25 | 0.75 | 0.55 | 1.39E-01 |
| 1.50 | 0.50 | 0.55 | 2.99E-01 |
| 1.75 | 0.25 | 0.55 | 5.05E-01 |
| 2.00 | 0.00 | 0.55 | 9.33E-01 |

Table 2: Breakdown voltage and Reverse current density (@500 V) for various $t_{ox}$.

| Field oxide thickness, $t_{ox}$ (μm) | Breakdown voltage (V) | *Reverse Current density @500 V* |
|---|---|---|
| 0.1 | 921 | 3.83E-05 |
| 0.2 | 933 | 5.07E-05 |
| 0.3 | 951 | 6.16E-05 |
| 0.4 | 982 | 7.15E-05 |
| 0.5 | 960 | 7.98E-05 |
| 0.6 | 948 | 8.72E-05 |
| 0.7 | 933 | 9.42E-05 |

Table 3: Optimized MEDICI input parameters for the 4H-SiC LDSS, LCS and LTSS rectifiers.

| Parameter | Value |
|---|---|
| N$^+$ doping for ohmic contact | $10^{20}$ cm$^{-3}$ |
| Drift region doping, $N_D$ | $1\times10^{17}$ cm$^{-3}$ |
| Drift region length, $L$ | 5.50 μm |
| Drift region thickness, $t$ | 2.00 μm |
| Field oxide thickness, $t_{ox}$ | 0.40 μm |
| Field plate length, $L_{FP}$ | 4.50 μm |
| Trench depth, $d_h$ | 1.75 μm |
| Trench depth, $d_l$ | 0.25 μm |
| Low Schottky barrier height (Ti), $\phi_{BL}$ | 0.85 eV |
| High Schottky barrier height (Ni), $\phi_{BH}$ | 1.50 eV |
| Richardson constant | 140 A/cm$^2$K$^{-2}$ |

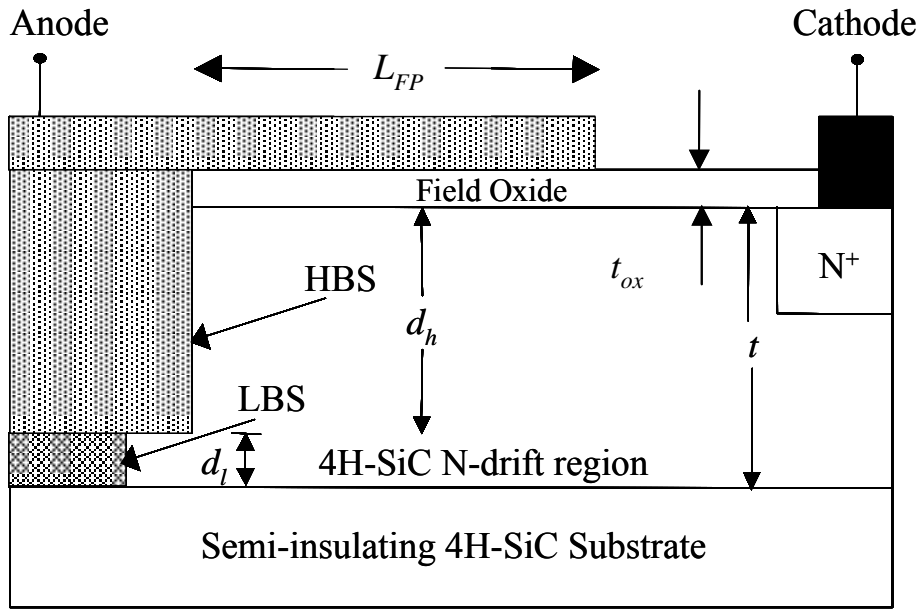

Fig. 1

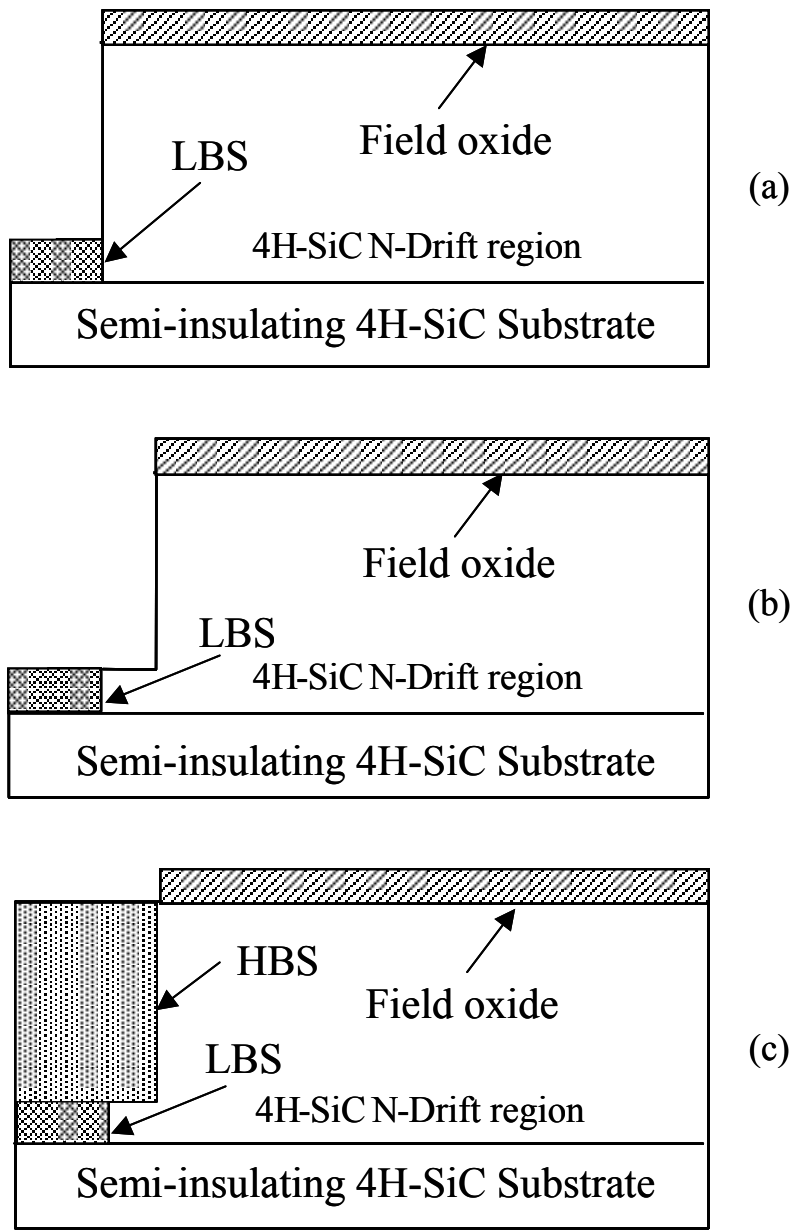

Fig. 2

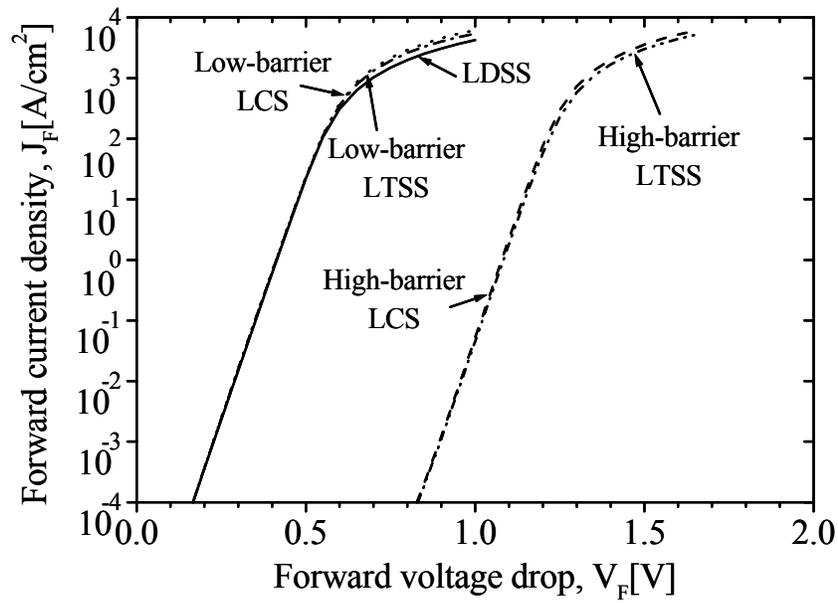

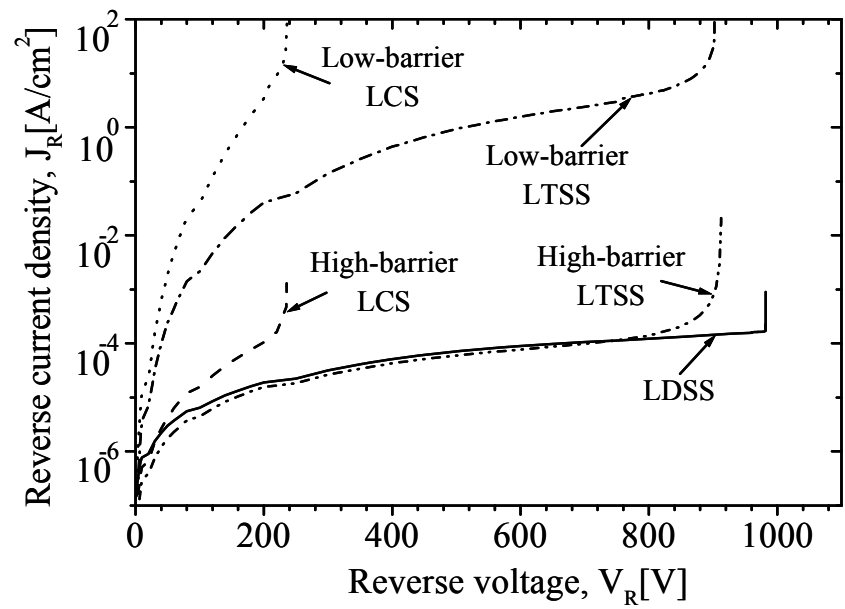

Fig. 3

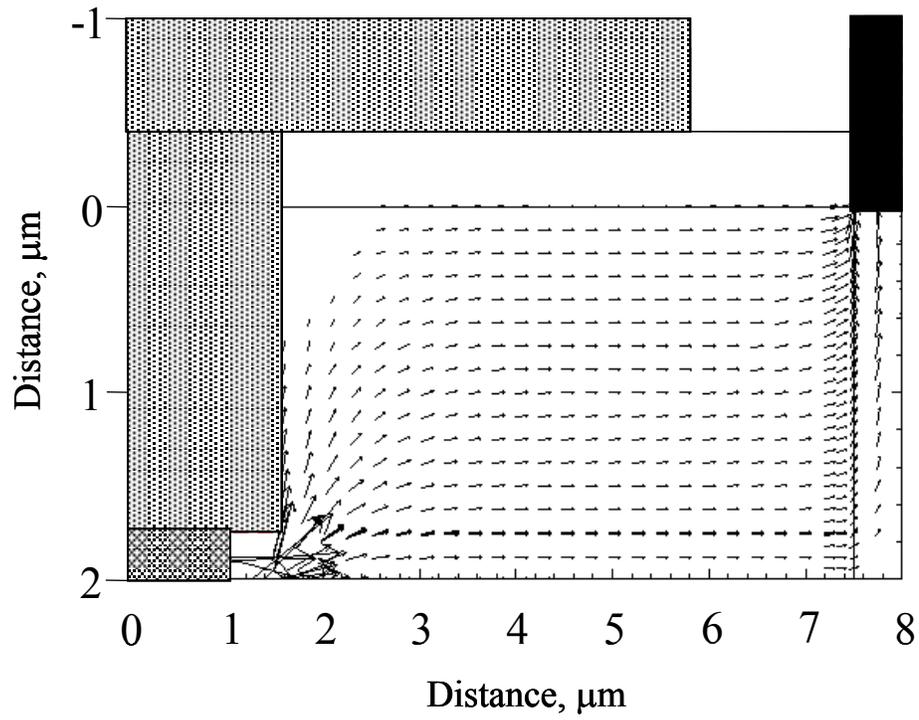

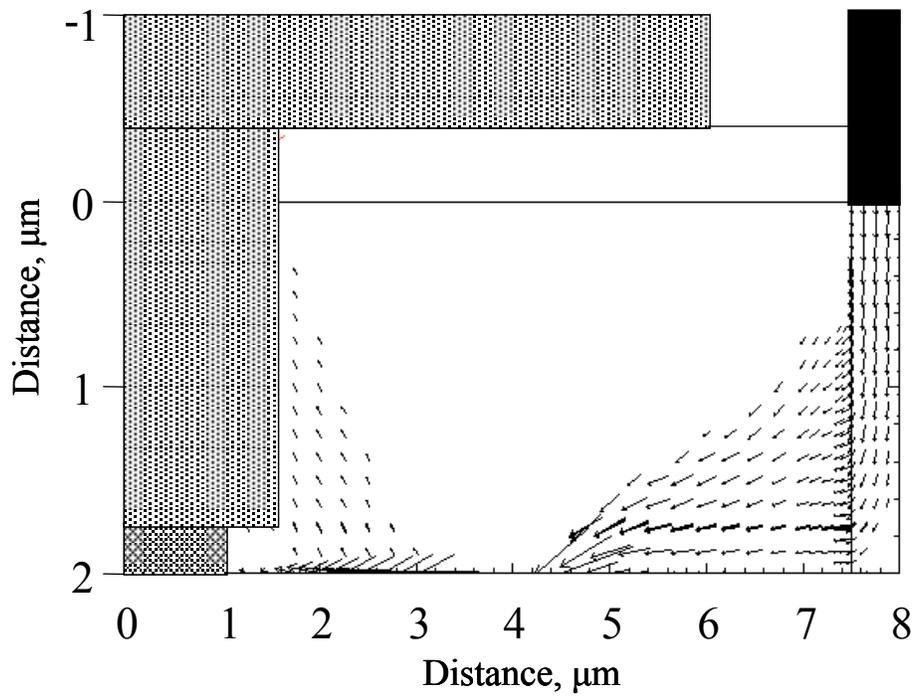

Fig. 4

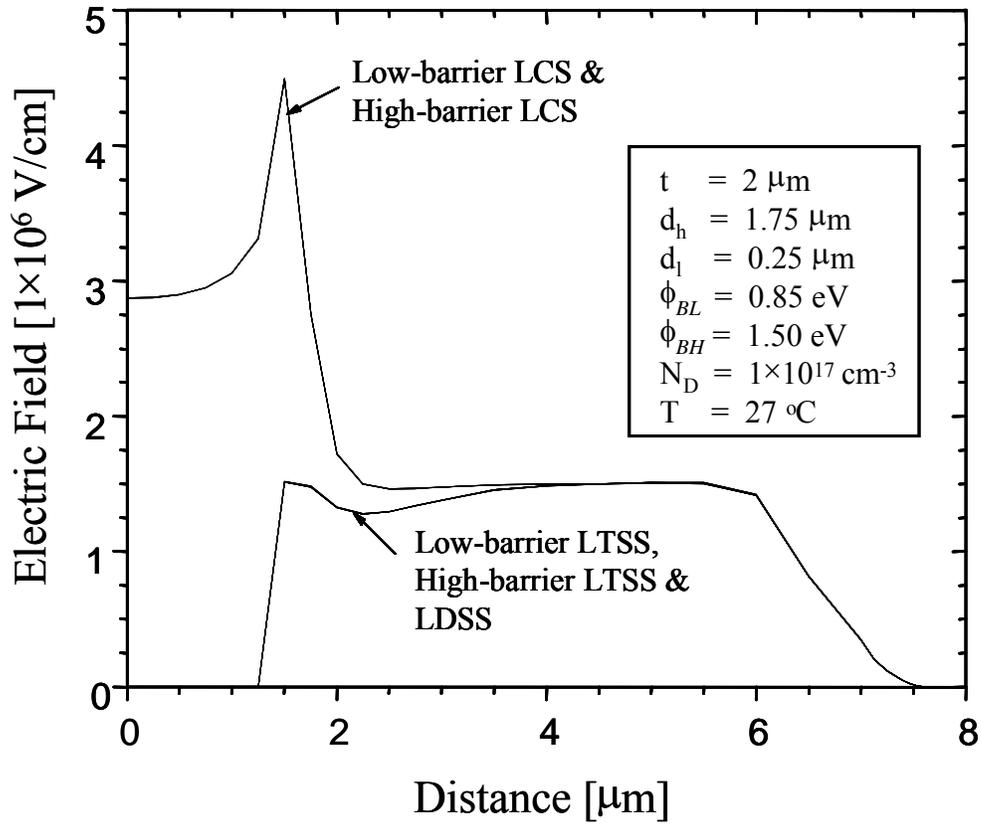

Fig. 5

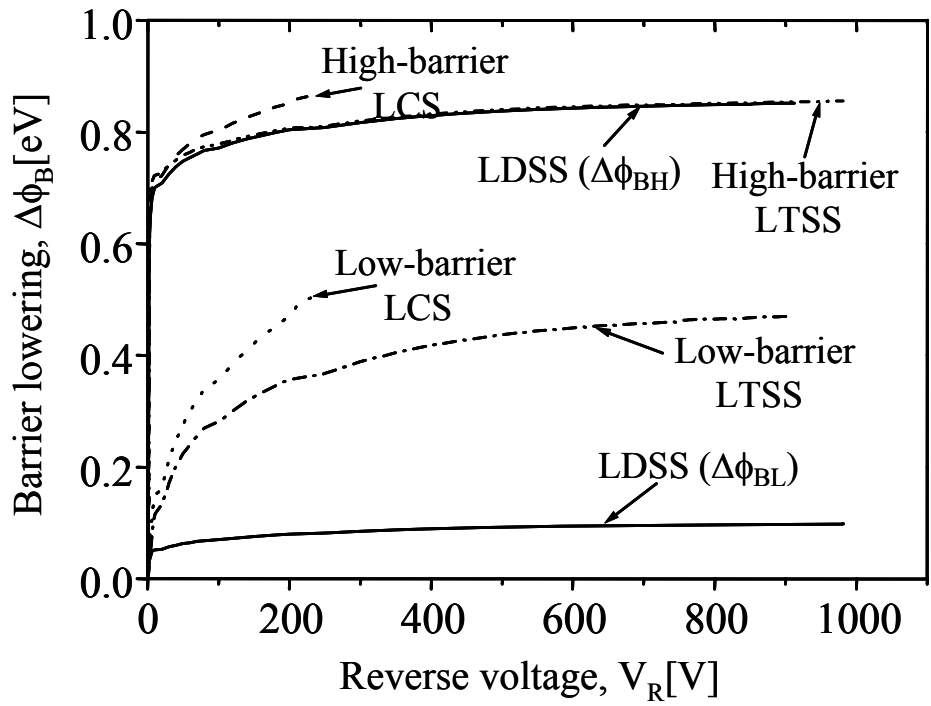

Fig. 6